\UseRawInputEncoding
\documentclass[a4paper,aps,prd,twocolumn,tightenlines,preprintnumbers,nofootinbib,showkeys, superscriptaddress]{revtex4-1}
\usepackage{fullpage}
\usepackage{amsfonts}
\usepackage{amsmath}
\usepackage{slashed}
\usepackage{amssymb}
\usepackage{graphicx}
\usepackage{makeidx}
\usepackage{cancel}
\usepackage{epic}
\usepackage{eepic}
\usepackage{epsfig}
\usepackage{latexsym}
\usepackage[dvipsnames]{xcolor}
\usepackage{float}
\usepackage{multirow}
\usepackage[export]{adjustbox}
\usepackage{xurl,hyperref}
\usepackage{enumitem}
\hypersetup{colorlinks=true,citecolor=red,linkcolor=NavyBlue,urlcolor=NavyBlue}
\usepackage[utf8]{inputenc}
\usepackage[caption=false]{subfig}
 
\usepackage{natbib}
\usepackage{relsize}
\usepackage[left=1.5 cm,right=1.5 cm,top=2 cm,bottom=2 cm]{geometry}
\usepackage{mathptmx}
\begin{document}
\relscale{1.05}
\captionsetup[subfigure]{labelformat=empty}

\title{Leptoquark-induced CLFV decays with a light SM-singlet scalar}

\author{Bibhabasu De}
\email{bibhabasude@gmail.com}
\affiliation{Department of Physics, The ICFAI University Tripura, Kamalghat-799210, India}

\date{\today}
%\preprint{IP-BBSR/2020-2}

\begin{abstract}
\noindent
The Standard Model~(SM), if augmented with a light SM-singlet scalar $\phi$ and a TeV-scale scalar leptoquark~(LQ) $S_1$, 2-body charged lepton flavor violating~(CLFV) decay channels can be accessed with $\phi$ as one of the final states, where the leading order effective interactions between $\phi$ and the SM fields arise at one-loop level. Further, in the presence of $S_1$, $\phi$ can mediate 3-body CLFV processes with either two photons or two gluons in the final states. Thus, the model predicts an exotic 3-body CLFV channel: $\ell_A\to\ell_B gg$, which can be tested/constrained only through some future high-energy experiments looking for di-gluon signals from a leptonic decay. 
\end{abstract}
	
\maketitle	

\section{Introduction}
\noindent
At the perturbative regime, Standard Model possesses an accidental symmetry represented by the global gauge group $\mathcal{G}_{LB}\supset U(1)_{B+L}\times U(1)_{B-L}\times U(1)_{L_\mu-L_\tau} \times U(1)_{L_\mu+L_\tau-2L_e}$~\cite{Heeck:2016xwg} which ensures the conservation of baryon number $B$ and the individual lepton numbers $L_A$~($A=e,\,\mu,\,\tau$). $L=\sum_{A=e,\,\mu,\,\tau}L_A$ denotes the total lepton number. However, at the non-perturbative scale, the $B+L$ part of the symmetry group is broken by 6 units~\cite{tHooft:1976rip}. At the same time, the neutrino oscillations strongly establish that the lepton flavor can't be a protected symmetry, resulting in a broken $U(1)_{L_\mu-L_\tau} \times U(1)_{L_\mu+L_\tau-2L_e}$ in the neutrino sector. The observation definitely motivates a search for similar flavor violations in the charged sector. However, for the charged leptons, $\mathcal{G}_{LB}$ is still a conserved symmetry as ensured by the Glashow-Iliopoulos-Maiani~(GIM) mechanism~\cite{Glashow:1970gm} --- any CLFV process induced by the non-zero neutrino masses and a non-trivial lepton mixing matrix will be too suppressed to detect. For example, the CLFV decay $\ell_A\to\ell_B\gamma$ induced through the Dirac neutrinos at one-loop level results in BR$(\ell_A\to\ell_B\gamma)<10^{-53}$  for all the possible channels~\cite{Petcov:1976ff}. Therefore, any observation of charged lepton flavor violation can be marked as a significant signature of Beyond Standard Model~(BSM) contribution, as well as physics beyond neutrino oscillations.

Experimental searches for CLFV processes have been going on for a long time~\cite{Hincks:1948vr, PhysRev.74.1364}, with the current sensitivity reaching $\sim\mathcal{O}(10^{-13})$ for a few flavor violating channels~\cite{MEG:2016leq, SINDRUM:1987nra, SINDRUMII:2006dvw} and being significantly improved for the others. However, in the absence of any positive signal, interest has grown in searching for CLFV decays involving low-energy BSM particles. For example, Belle-II presents the best current upper limit on $\tau\to e\phi$ and $\tau\to \mu\phi$ at 95\% C.L.~\cite{Belle-II:2022heu}, where $\phi$ is an invisible spin-0 boson in the mass range $0-1.6$ GeV. Similar bounds for $\mu\to e \phi$ are given by Refs.~\cite{TWIST:2014ymv,Jodidio:1986mz}. However, if $\phi$ decays to visible particles inside the detector, more stringent constraints could be placed from the non-observation of 3-body CLFV processes. 

Various models have been proposed to address the current bounds on $\ell_A\to\ell_B\phi$, with $\phi$ being either a light scalar/pseudoscalar~\cite{PhysRevLett.48.11, Wilczek:1982rv,Grinstein:1985rt, Berezhiani:1989fp, Feng:1997tn, Hirsch:2009ee,Jaeckel:2013uva,Celis:2014iua,Celis:2014jua, Galon:2016bka,Calibbi:2016hwq,Ema:2016ops,Bjorkeroth:2018dzu,Bauer:2019gfk,Heeck:2019guh, Cornella:2019uxs, Calibbi:2020jvd,Escribano:2020wua} or a gauge boson~\cite{Farzan:2015hkd, Heeck:2016xkh,Farzan:2016wym,Ibarra:2021xyk}, associated with some spontaneously broken symmetry. This paper considers $\phi$ to be a generic real SM-singlet scalar where the flavor-violating interactions arise at one-loop level in the presence of a scalar LQ $S_1$. LQs~(for a review, see Ref.~\cite{Dorsner:2016wpm}) appear naturally in the grand unified theories~(GUT)~\cite{Pati:1974yy, Georgi:1974sy, Georgi:1974my} and work as an excellent BSM candidate to induce flavor violation. Though in the simplest GUT models LQs acquire mass at a scale $\Lambda_{\rm GUT}\gg \mathcal{O}(1)$ TeV, there exist GUT formulations that can ensure the stability of proton with a TeV-scale scalar LQ~\cite{Dorsner:2004jj, Aydemir:2019ynb, Dorsner:2024seb}. Thus, the present paper, while extending the SM with a TeV-scale scalar LQ to describe the interactions between $\phi$ and the SM fields, effectively portrays a UV-complete theory. Ref.~\cite{Mandal:2019gff} has already elaborated on various CLFV observables arising in the presence of a TeV-scale scalar LQ and used the experimental upper limits to constrain the New Physics~(NP) couplings. However, interesting consequences may appear if one considers a light SM-singlet scalar within the $S_1$-extension of SM. For example, with $M_\phi<(m_\mu-m_e)$, $\phi$ can be produced through the on-shell decays of $\tau$ and $\mu$ leading to the CLFV processes $\tau\to\ell_B\phi$~[$\ell_B=e,\,\mu$] and $\mu\to e \phi$, respectively. In practice, the term {\it light} will be used in this paper to refer to $M_\phi<2m_e$, i.e., a sub-MeV gauge-singlet scalar. In this mass regime, it can be easily ensured that the di-lepton and di-quark decay modes of $\phi$~\cite{De:2024tbo} are kinematically forbidden. However, $\phi$ can always decay to two photons at one-loop level, leading to 3-body CLFV channels of the form $\ell_A\to\ell_B\gamma\gamma$. Thus, the experimental bounds for $\ell_A\to\ell_B\gamma\gamma$ can be used to check the consistency of the parameter space allowed through the $\ell_A\to\ell_B\phi$ searches. Further, LQs being colored particles, within this proposed BSM formulation $\phi$ can also decay to two gluons. This results in a rarely discussed unconstrained CLFV decay process: $\ell_A\to\ell_Bgg$. Indeed, this is a significant phenomenological outcome of this model as it predicts a set of unexplored CLFV observables. Therefore, any future search aiming for $\ell_A\to\ell_Bgg$ will be crucial to test/falsify the model.

The rest of the paper is arranged as follows. Sec.~\ref{sec:1} describes the NP interactions at the TeV scale, which connect $\phi$ with the SM fields. The analytical results for $\ell_A\to\ell_B\phi$ have been established in Sec.~\ref{sec:2}. Sec.~\ref{sec:3} presents a parameter space consistent with all the existing CLFV constraints that can appear in this chosen framework and, hence, numerically analyze the detection prospects of $\ell_A\to\ell_B\phi$ within the considered mass regime. The possible decay modes of $\phi$ and the corresponding $\phi$-mediated 3-body CLFV processes have been discussed in Sec.~\ref{sec:4}. Finally, the work has been concluded in Sec.~\ref{sec:5}. 
\section{The Model: A Simple Extension of the SM}
\label{sec:1}
\noindent
The model considers a GUT-motivated minimal extension of the SM where NP interactions arise at a scale $\Lambda_{\rm NP}\sim\mathcal{O}(1)$ TeV. The augmented particle spectrum includes a scalar LQ $S_1$ which transforms as ($\bar{\mathbf{3}}$, {\bf 1}, 1/3) under the SM gauge group $\mathcal{G}_{\rm SM}\supset SU(3)_C\times SU(2)_L\times U(1)_Y$. Further, a real SM-singlet scalar $\phi$ can be proposed to exist in the sub-MeV mass regime such that its decays to the SM fields are either kinematically forbidden or highly suppressed. Defining the electromagnetic~(EM) charge as $Q_{\rm EM}=T_3+Y$, the complete particle spectrum of the present framework has been listed in Table~\ref{tab:parti}.
\begin{table}[!ht]
\begin{tabular}{|c|c|c|}
\hline
Fields & Generations & $SU(3)_C\times SU(2)_L\times U(1)_Y$ \\
\hline
\hline
$L_L=(\nu_L\quad \ell_L)^T$  & 3 & ({\bf 1}, {\bf 2}, -1/2)  \\
		$\ell_R= (e_R,\,\mu_R,\,\tau_R)$ & 3 & ({\bf 1}, {\bf 1}, -1) \\
		$Q_L=(u_L\quad d_L)^T$  & 3 & ({\bf 3}, {\bf 2}, 1/6) \\
		$U_R=(u_R,\,c_R,\,t_R)$  & 3 & ({\bf 3}, {\bf 1}, 2/3) \\
		$D_R=(d_R,\,s_R,\,b_R)$  & 3 & ({\bf 3}, {\bf 1}, -1/3) \\
		$H = (H^+ \quad H^0)^T $ & 1 & ({\bf 1}, {\bf 2}, 1/2)  \\
\hline
		$S_1$ & 1 & ($\bar{\mathbf{3}}$, {\bf 1}, 1/3)	\\
		$\phi$ & 1 & ({\bf 1}, {\bf 1}, 0)	\\
		\hline
\end{tabular} 
\caption{Fields and their transformations under $\mathcal{G}_{\rm SM}$.}
\label{tab:parti}
\end{table}

The NP interactions can be described through
\begin{align}
\mathcal{L}_{\rm NP}=~& (\mathcal{D}^\mu S_1)^\dagger(\mathcal{D}_\mu S_1)+\frac{1}{2}(\partial^\mu\phi)(\partial_\mu\phi) \nonumber\\
&-\Big[Y_L^{ij} (\bar{Q}_L^{Cia}\epsilon^{ab}L_L^{jb})S_1+ Y_R^{ij} 
(\bar{U}_R^{Ci}\ell_R^j)S_1+{\rm h.c.}\Big]\nonumber\\
&\qquad\qquad\qquad\qquad\qquad\qquad-\mathbb{V}(H,\,S_1,\,\phi),
\label{eq:L_NP}
\end{align}
where $Y_{L}$ and $Y_R$ are completely arbitrary $3\times 3$ Yukawa matrices in the flavor basis. The superscript $C$ defines the charge-conjugated states; $\{a,b\}$ and $\{i,j\}$ stand for the $SU(2)_L$ and flavor indices, respectively. $\epsilon^{ab}=(i\sigma_2)^{ab}$, $\sigma_k$~($k=1,\,2,\,3$) being the Pauli matrices. The color indices have been suppressed for simplicity. Note that, the gauge interactions of $S_1$ can easily be obtained with an explicit definition of the covariant derivative $\mathcal{D}^\mu$. These interactions can be useful to produce $\phi$ at the colliders~\cite{Bhaskar:2020kdr, De:2024tbo}.

The freedom to rotate equal-isospin fermion fields in the flavor basis, permits one to assume the charged lepton and down-type quark Yukawas to be diagonal so that the transformation from flavor to mass basis is given by $u_L\to \left(\mathbf{V}_{\rm CKM}^\dagger\right) u_L$ and $\nu_L\to \left(\mathbf{U}_{\rm PMNS}\right) \nu_L$. Here $\mathbf{V}_{\rm CKM}$ and $\mathbf{U}_{\rm PMNS}$ represent the CKM and PMNS matrices, respectively. Therefore, in the physical basis,   
\begin{align}
\mathcal{L}_{\rm NP}&= (\mathcal{D}^\mu S_1)^\dagger(\mathcal{D}_\mu S_1)+\frac{1}{2}(\partial^\mu\phi)(\partial_\mu\phi)\nonumber\\
&-\Big[\left\{\bar{u}_L^{Ci}\left(\mathbf{V}^*_{\rm CKM}Y_L\right)^{ij} \ell_L^{j}\right\}S_1
-\left\{\bar{d}_L^{Ci}\left(Y_L \mathbf{U}_{\rm PMNS}\right)^{ij} \nu_L^{j}\right\}S_1\nonumber\\
&\qquad\qquad+ Y_R^{ij} 
(\bar{U}_R^{Ci}\ell_R^j)S_1+{\rm h.c.}\Bigg]-\mathbb{V}(H,\,S_1,\,\phi)\nonumber\\
&= (\mathcal{D}^\mu S_1)^\dagger(\mathcal{D}_\mu S_1)+\frac{1}{2}(\partial^\mu\phi)(\partial_\mu\phi)-\Big[\left(\bar{u}_L^{Ci}\xi_L^{ij} \ell_L^{j}\right)S_1\nonumber\\
&-\left(\bar{d}_L^{Ci}\chi^{ij} \nu_L^{j}\right)S_1+ \xi_R^{ij} 
(\bar{U}_R^{Ci}\ell_R^j)S_1+{\rm h.c.}\Bigg]\nonumber\\
&\qquad\qquad\qquad\qquad\qquad\qquad-\mathbb{V}(H,\,S_1,\,\phi),
\label{eq:NP_Lag}
\end{align}
where, $\xi_L^{ij}=\left(\mathbf{V}^*_{\rm CKM}Y_L\right)^{ij}$, $\chi^{ij}=\left(Y_L \mathbf{U}_{\rm PMNS}\right)^{ij}$, and $\xi_R^{ij}=Y_R^{ij}$. However, the neutrino sector doesn't play any role in the present analysis. The scalar potential can be cast as,
	\begin{align}
\mathbb{V}&(H,\, S_1,\,\phi)= \lambda_{HS_1} \left(H^\dagger H\right) \left(S^\dag_1S_1\right)+\beta\phi 
\left(S^\dag_1S_1\right)+\delta\phi(H^\dagger H)\nonumber\\
& +\lambda_{H\phi} (H^\dagger H)\phi^2+\lambda_{\phi S_1}\phi^2 
\left(S^\dag_1S_1\right)+ \frac{1}{2}M^2_\phi \phi^2 + M^{2}_{S_1}\left(S^\dag_1S_1\right).
	\label{eq:pot}
	\end{align}
However, one can assume $\lambda_{H\phi}$ and $\delta$ to be negligible so that the presence of $\phi$ doesn't affect the SM scalar sector. Further, by setting $\lambda_{HS_1}\to 0$, one can ensure that the trilinear Higgs coupling assumes its SM-predicted value and doesn't introduce any additional constraint to the parameter space. Note that, $\beta$ is a mass dimensional coupling which should, in principle, represent the highest scale of the theory and hence must be $\sim\mathcal{O}(1)$ TeV. After electroweak symmetry breaking~(EWSB), the physical masses can be obtained via,
	 \begin{align}
H = 
\frac{1}{\sqrt{2}}\begin{pmatrix}
0 \\
v + h
\end{pmatrix}, \quad\quad \phi=\phi,
\end{align}
where $h$ denotes the physical Higgs field, and $v=246$ GeV defines the electroweak vacuum expectation value~(VEV). However, in the limit $\lambda_{HS_1}\to 0$ and $\lambda_{H\phi}\to 0$, $M_{S_1}$ and $M_\phi$ represent the physical masses of $S_1$ and $\phi$, respectively.
\section{Lepton Flavor Violation: $\pmb{\Large \ell_A\to \ell_B+\phi}$}
\label{sec:2}
\begin{figure}[!ht]
   \centering
   \includegraphics[scale=0.5]{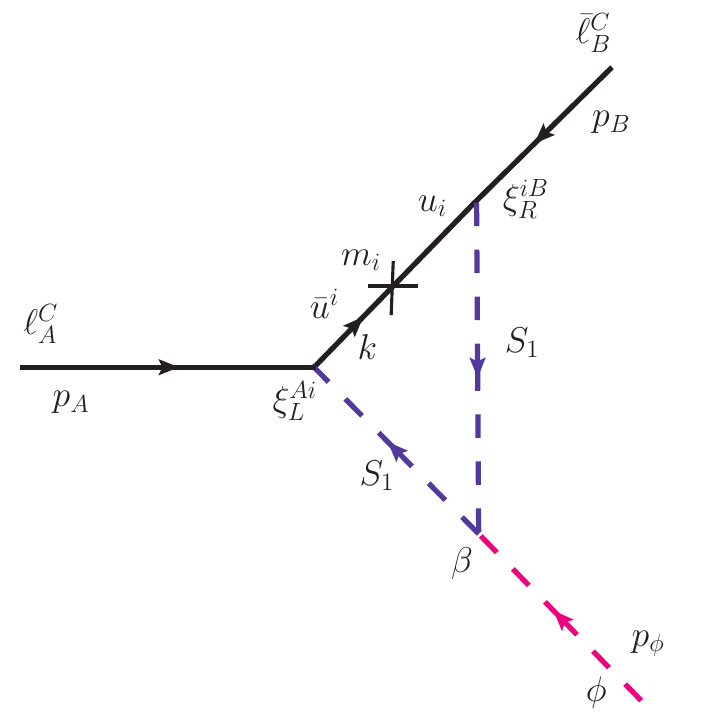}
   \caption{Leading order contribution to $\ell_A\to \ell_B\phi$. Here, $k$ denotes the loop momentum, and $p_A=-(p_B+p_\phi)$ defines the momentum conservation.}
   \label{fig:tau_lep}
   \end{figure}

\noindent
The proposed model, which extends the SM with a scalar LQ $S_1$, possesses a rich BSM phenomenology in the lepton sector. For example, the discrepancies in muon and electron $(g-2)$, semi-leptonic decays, non-observation of CLFV processes including $\ell_A\to\ell_B\gamma$, $\ell_A\to \bar{\ell}_B\ell_B\ell_B$, $h\to \bar{\ell}_A\ell_B$, and $\mu-e$ conversion can be explained with $M_{S_1}\sim\mathcal{O}(1)$ TeV. However, in the presence of an additional light SM-singlet scalar, new CLFV channels can be introduced within the considered framework where the SM leptons~($\tau$ and $\mu$) can decay to $\phi$ and a lighter lepton through the effective vertex shown in Fig.~\ref{fig:tau_lep}. To be specific, for $M_\phi\leq 1$ MeV, $\tau\to e\phi$, $\tau\to \mu\phi$, and $\mu\to e\phi$ CLFV modes are kinematically accessible.   
   
The effective coupling corresponding to Fig.~\ref{fig:tau_lep} is given by,
\begin{align}
-i&Y_{\ell_A\ell_B\phi}= \sum_{i=\,u,c,t}(\xi_L^{A i})^*\xi_R^{iB}\beta m_i\int\frac{d^4k}{(2\pi)^4} \Bigg[\frac{(k^2+m_i^2)}{(k^2-m_i^2)^2}\times\nonumber\\
&\qquad\qquad\frac{1}{\{(k+p_B)^2-M_{S_1}^2\}\{(k+p_B+p_\phi)^2-M_{S_1}^2\}}\Bigg]\nonumber\\
=& \sum_{i=\,u,c,t}(\xi_L^{A i})^*\xi_R^{iB}\beta m_i\int\frac{d^4k}{(2\pi)^4} \Bigg[\nonumber\\
&\frac{1}{(k^2-m_i^2)\{(k+p_B)^2-M_{S_1}^2\}\{(k+p_B+p_\phi)^2-M_{S_1}^2\}}\nonumber\\
&+\frac{2m_i^2}{(k^2-m_i^2)^2\{(k+p_B)^2-M_{S_1}^2\}\{(k+p_B+p_\phi)^2-M_{S_1}^2\}}\Bigg]\nonumber\\
=& \left(\frac{i\,\beta}{16\pi^2}\right)\sum_{i=\,u,c,t}\Omega^{AB}_i\, m_i\,\Bigg[\mathcal{J}_1+2m_i^2\times \mathcal{J}_2\Bigg],
\label{eq:Y1}
\end{align}
where $\Omega^{AB}_i=(\xi_L^{A i})^*\xi_R^{iB}$, and
\begin{align}
\mathcal{J}_1=&\ -C_0\left(m_B^2,\,M_\phi^2,\,m_A^2,\, m_i^2, \,M_{S_1}^2,\,M_{S_1}^2\right),\nonumber\\
\mathcal{J}_2=&\ D_0\left(0,\,m_B^2,\,M_\phi^2,\,m_A^2,\,m_B^2,\,m_A^2,\, m_i^2,\,m_i^2,\, M_{S_1}^2,\,M_{S_1}^2\right).
\end{align}
Eq.~\eqref{eq:Y1} has been obtained assuming an on-shell decay of $\ell_A$ so that $p_A^2=(p_B+p_\phi)^2=m_A^2$. Here, $C_0$ and $D_0$ denote the standard Passarino-Veltman functions~\cite{Romao:2020,Hahn:1998yk} for the scalar 3-point and 4-point one-loop integrals, respectively.

In the rest-frame of $\ell_A$, the decay width for $\ell_A\to\ell_B\phi$ can be calculated as,
\begin{align}
\Gamma(\ell_A\to\ell_B\phi)=&~\frac{\mathcal{N}_C\times|Y_{\ell_A\ell_B\phi}|^2}{16\pi}\times m_A\times\Big[(1+\rho_B)^2-\rho_\phi^2\Big]\nonumber\\
&\times\Bigg[\Big\{1-(\rho_\phi+\rho_B)^2\Big\}\Big\{1-(\rho_\phi-\rho_B)^2\Big\}\Bigg]^{1/2}\, ,
\label{eq:dec}
\end{align}
where $\rho_B=m_B/m_A$, and $\rho_\phi=M_\phi/m_A$. $\mathcal{N}_C=3$ stands for the color degeneracy factor. However, to detect any signal from $\ell_A\to\ell_B \phi$ decay, $\ell_A\to\ell_B\bar{\nu}_B\nu_A$ acts as the strongest background. Therefore, to parametrize the detectability of $\ell_A\to\ell_B\phi$ over $\ell_A\to\ell_B\bar{\nu}_B\nu_A$, a variable $R_{AB}$ can be defined as,
\begin{align}
R_{AB}=\frac{{\rm BR}(\ell_A\to\ell_B\phi)}{{\rm BR}(\ell_A\to\ell_B\bar{\nu}_B\nu_A)},
\label{eq:R}
\end{align}
where BR symbolizes the branching ratio for a particular decay mode. In the subsequent analysis, $R_{AB}$ will be considered as a {\it good} observable to constrain the parameter space. 
\section{Numerical Analysis}
\label{sec:3}
\noindent
As stated in Sec.~\ref{sec:2}, the $S_1$-extension of SM can accommodate several BSM observations and possibilities resulting in strong bounds on the NP couplings $\xi^{ij}_{L,R}$. Thus, to analyze the $\ell_A\to\ell_B\phi$ processes, one must choose a parameter space consistent with all the existing leptonic constraints. Following Ref.~\cite{Mandal:2019gff}, Table~\ref{tab:bound} enlists the most stringent upper limits on $|\Omega^{AB}_i|^2$ relevant for the present study. 
\begin{table}[!ht]
\begin{tabular}{|c|c|c|c|}
\hline
CLFV & $|\Omega^{AB}_i|^2$ & Experimental & Considered\\
Processes & & Bounds $\times\left(M_{S_1}/{\rm TeV}\right)^4$ & Values\\
\hline
\hline
 & $|\Omega^{\tau e}_u|^2$ & $<\,3.0\times 10^{-6}$~\cite{ParticleDataGroup:2018ovx} & $10^{-6}$\\
$\tau\to e\phi$ & $|\Omega^{\tau e}_c|^2$ & $<\,2.1\times 10^{-4}$~\cite{BaBar:2009hkt} & $10^{-6}$\\
 & $|\Omega^{\tau e}_t|^2$ & $<\,3.8\times 10^{-7}$~\cite{BaBar:2009hkt} & $[10^{-12},\,10^{-6}]$\\
\hline
 & $|\Omega^{\tau \mu}_u|^2$ & $<\,5.7\times 10^{-6}$~\cite{ParticleDataGroup:2018ovx} & $10^{-6}$\\
$\tau\to \mu\phi$ & $|\Omega^{\tau \mu}_c|^2$ & $<\,2.8\times 10^{-4}$~\cite{BaBar:2009hkt} & $10^{-6}$\\
 & $|\Omega^{\tau \mu}_t|^2$ & $<\,5.0\times 10^{-7}$~\cite{BaBar:2009hkt} & $[10^{-12},\,10^{-6}]$\\
 \hline
 & $|\Omega^{\mu e}_u|^2$ & $<\,6.7\times 10^{-12}$~\cite{SINDRUMII:2006dvw} & $10^{-12}$\\
$\mu\to e\phi$ & $|\Omega^{\mu e}_c|^2$ & $<\,1.7\times 10^{-12}$~\cite{MEG:2016leq} & $[10^{-16},\,10^{-12}]$\\
 & $|\Omega^{\mu e}_t|^2$ & $<\,3.0\times 10^{-15}$~\cite{MEG:2016leq} & $10^{-17}$\\
 \hline
\end{tabular} 
\caption{Experimental constraints on the NP couplings and their considered values used for the computations.}
\label{tab:bound}
\end{table}
The first column of Table~\ref{tab:bound} specifies a particular $\ell_A\to\ell_B\phi$ process, while the third column represents the experimental bounds on the associated NP couplings arising through the other CLFV processes. However, the bounds drastically deteriorate with increasing LQ mass. In general, the couplings considered for the present analysis assume $M_{S_1}\geq 2$ TeV~\cite{ParticleDataGroup:2022pth} and have been listed in the last column of Table~\ref{tab:bound}. As an illustrative choice, $\beta=5$ TeV will be used for the analysis.
\subsection{$\pmb{\tau}$-Sector}
\begin{figure}[!ht]
   \centering
   \includegraphics[scale=0.65]{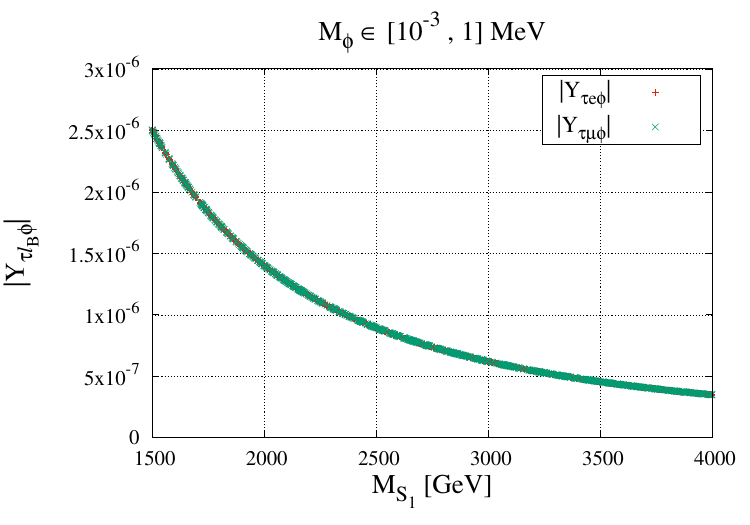}
   \caption{Variation of $|Y_{\tau e\phi}|$~(red) and $|Y_{\tau \mu\phi}|$~(green) as a function of $M_{S_1}$ with $M_\phi\in[1~{\rm keV},1~{\rm MeV}]$.}
   \label{fig:Yukawa1}
   \end{figure}

The $\tau e\phi$ and $\tau \mu\phi$ effective couplings can be obtained from Eq.~\eqref{eq:Y1} with $m_A=m_\tau$ and $m_B=m_e,\,m_\mu$, respectively. Fig.~\ref{fig:Yukawa1} shows their variations with increasing LQ mass. All the NP couplings $\Omega^{\tau B}_i$ have been fixed at their respective upper limits, i.e., $10^{-3}$, while $M_\phi$ has been randomly varied between 1 keV and 1 MeV to generate Fig.~\ref{fig:Yukawa1}. 
However, the results suggest that the variation of $M_\phi$ in the considered range negligibly affects $|Y_{\tau\ell_B\phi}|$~($\ell_B=e,\,\mu$). Further, the lighter lepton masses~(i.e., $m_e$ and $m_\mu$) being ignorable with respect to $\Lambda_{\rm NP}$, $|Y_{\tau e\phi}|$ and $|Y_{\tau\mu\phi}|$ overlap.
   \begin{figure}[!ht]
   \centering
   \subfloat[\qquad\qquad(a)]{\includegraphics[scale=0.65]{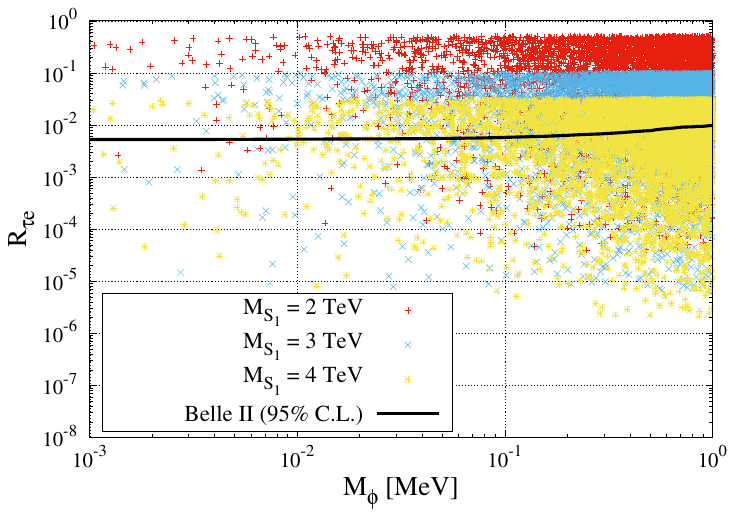}}\\
   \subfloat[\qquad\qquad(b)]{\includegraphics[scale=0.65]{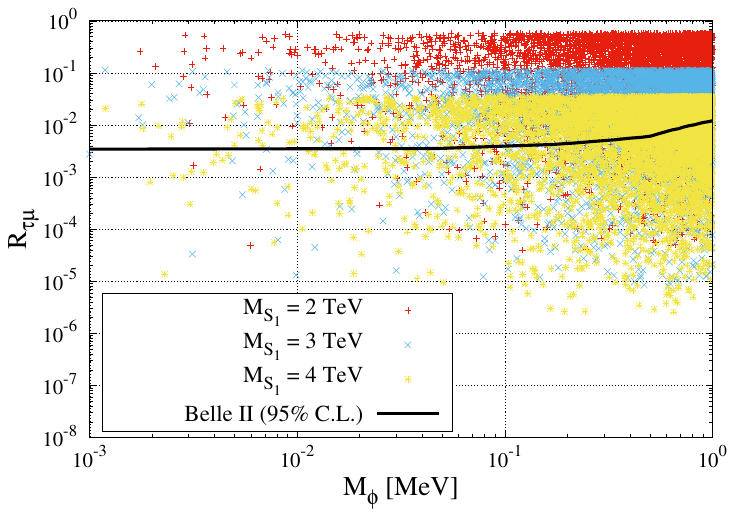}}
   \caption{Variation of (a) $R_{\tau e}$, and (b) $R_{\tau\mu}$ as a function of $M_\phi$ for $M_{S_1}=2$ TeV~(red), 3 TeV~(sky), and 4 TeV~(yellow). The solid black lines denote the observed upper limits~(at 95\% C.L.) from Belle II collaboration~\cite{Belle-II:2022heu}.}
   \label{fig:R1}
   \end{figure}
Fig.~\ref{fig:R1}\,(a) and \ref{fig:R1}\,(b) display the variation of $R_{\tau e}$ and $R_{\tau \mu}$ as a function of $M_\phi$, respectively. For Fig.~\ref{fig:R1}\,(a), $\Omega^{\tau e}_t$ has been randomly varied within the range $[10^{-6},\,10^{-3}]$, whereas one has to vary $\Omega^{\tau \mu}_t$ to generate the scattered plot in Fig.~\ref{fig:R1}\,(b). Currently, the most stringent bounds on $R_{\tau\ell_B}$ are set by Belle II collaboration~\cite{Belle-II:2022heu} as shown by the solid black lines in Fig.~\ref{fig:R1}. The results indicate a significant parameter space where the non-observation of $\tau\to e\phi$ and $\tau\to\mu\phi$ can be explained along with the other CLFV processes. Moreover, the allowed parameter space improves with a heavier $S_1$.
 \subsection{$\pmb{\mu}$-Sector}
\begin{figure}[!ht]
   \centering
   \includegraphics[scale=0.65]{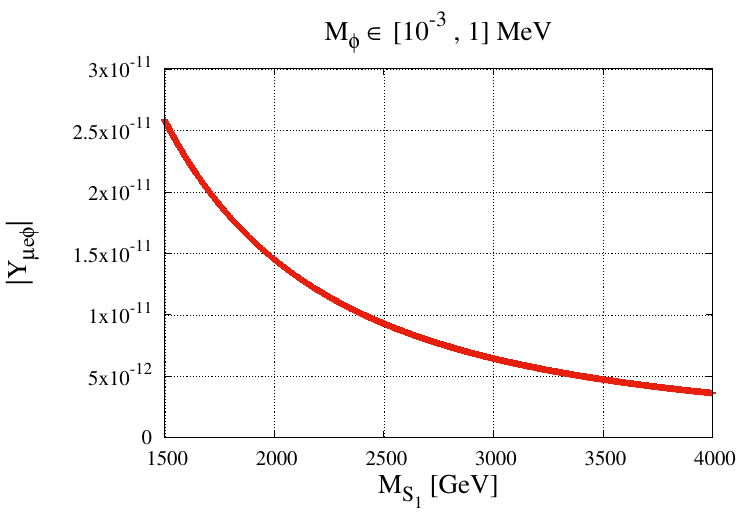}
   \caption{Variation of $|Y_{\mu e\phi}|$ as a function of $M_{S_1}$ with $M_\phi\in[1~{\rm keV},1~{\rm MeV}]$.}
   \label{fig:Yukawa}
   \end{figure}

$\phi$ being in the sub-MeV mass regime, $\mu$ can also decay to $\phi$ and $e$ through the  effective $Y_{\mu e \phi}$ coupling. Fig.~\ref{fig:Yukawa} depicts its variation as a function of $M_{S_1}$. As before, all the NP couplings have been fixed at their allowed upper bounds with $M_\phi$ varying randomly within the range 1 keV to 1 MeV.  
Note that, $|Y_{\mu e \phi}/Y_{\tau \ell_B \phi}|\sim 10^{-5}$. The large suppression for $Y_{\mu e \phi}$ stems from the NP couplings $\Omega^{\mu e}_i$~($i=u,\,c,\,t$), which are tightly constrained through the experimental bounds on $\mu\to e\gamma$~\cite{MEG:2016leq} and $\mu-e$ conversion on gold nuclei~\cite{SINDRUMII:2006dvw}.
   \begin{figure}[!ht]
   \centering
   \includegraphics[scale=0.65]{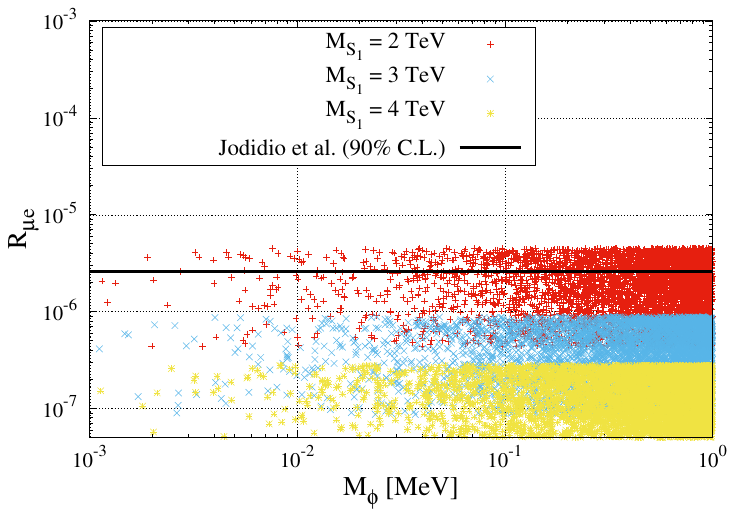}
   \caption{Variation of $R_{\mu e}$ as a function of $M_\phi$ for $M_{S_1}=2$ TeV~(red), 3 TeV~(sky), and 4 TeV~(yellow). The black solid line marks the observed upper limit~(at 90\% C.L.) from Jodidio et al.~\cite{Jodidio:1986mz}.}
   \label{fig:R}
   \end{figure}
   
Fig.~\ref{fig:R} presents the $M_\phi$ dependence of $R_{\mu e}$ for three different LQ masses. The plot has been generated through a random variation of $\Omega^{\mu e}_c$ within the range $[10^{-8},\,10^{-6}]$, whereas $\Omega^{\mu e}_u$ and $\Omega^{\mu e}_t$ have been fixed at $10^{-6}$ and $3.16\times 10^{-9}$, respectively. 

TWIST collaboration sets the most stringent bound on $R_{\mu e}$, when $\phi$ is a massive invisible boson in the range $13-80$ MeV~\cite{TWIST:2014ymv}. However, in the sub-MeV mass regime, $\phi$ can be treated as effectively massless, and hence, the best experimental upper limit can be read as $R_{\mu e}<2.6\times  10^{-6}$~(Jodidio et al.~\cite{Jodidio:1986mz}). Thus, in Fig.~\ref{fig:R}, the portion below the black solid line defines the allowed parameter space for $\mu\to e\phi$.

\section{Decay of $\pmb{\phi}$ and Other CLFV Constraints}
\label{sec:4}
\noindent
As indicated in the Introduction, Refs.~\cite{Jodidio:1986mz,TWIST:2014ymv,Belle-II:2022heu} set the best constraints for the considered 2-body CLFV decays if $\phi$ is an invisible light boson, i.e., either $\phi$ decays to invisible particles~(e.g., dark matter, neutrinos) or features a significantly large decay-length. In the former case, only a sharp energy peak of electron or muon can be searched over the continuous Michel spectrum. However, if $\phi$ decays to visible particles, there is always a chance to have better constraints from the 3-body CLFV processes. 

\begin{figure}[!ht]
\centering
\includegraphics[scale=0.37]{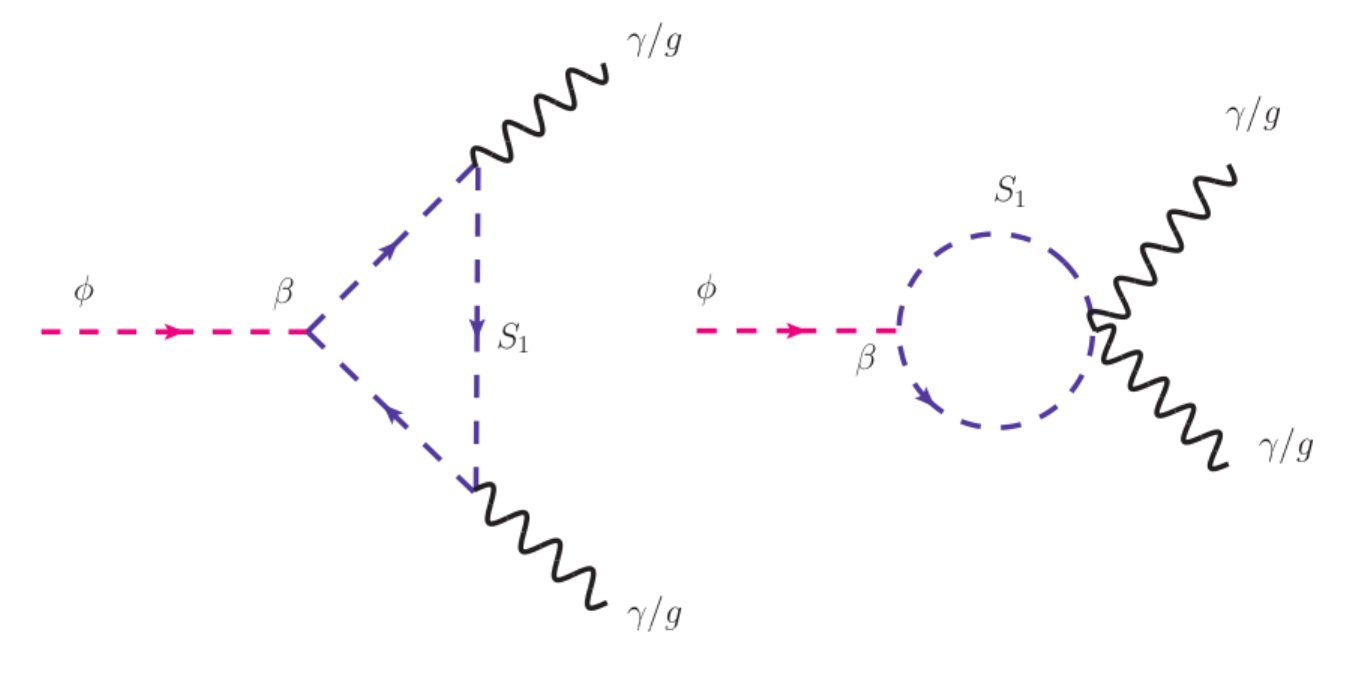}
\caption{Leading order contributions to $\phi\to \gamma\gamma\,(gg)$ processes.}
\label{fig:phi_gg}
\end{figure}

In the proposed framework, the gauge-singlet scalar $\phi$ produced through the on-shell decays of $\tau$ and/or $\mu$ can further decay via di-photon and di-gluon channels with the leading order contributions arising at one-loop level~[see Fig.~\ref{fig:phi_gg}]. Note that, for $M_\phi<2m_e$, other decay modes of $\phi$ are kinematically forbidden.

Thus, the total decay width of $\phi$ can be formulated as,
\begin{align}
\Gamma_\phi=\Gamma(\phi\to gg)+\Gamma(\phi\to\gamma\gamma),
\end{align}
where the partial decay widths are given by~\cite{Dorsner:2016wpm, Bhaskar:2020kdr},
\begin{align}
\Gamma(\phi\rightarrow gg)&\ =\ \frac{G_{\rm F}\alpha_S^2M_\phi^3}{64\sqrt{2}\pi^3}\left|\frac{\beta v }{2M_{S_1}^2}\mathcal{P}\left(\frac{M_\phi^2}{4M_{S_1}^2}\right)\right|^2, \nonumber\\
\Gamma(\phi\rightarrow\gamma\gamma)&\ =\ \frac{G_{\rm F}\alpha_{\rm EM}^2M_\phi^3}{128\sqrt{2}\pi^3}\left|\frac{\beta v }{6M_{S_1}^2}\mathcal{P}\left(\frac{M_\phi^2}{4M_{S_1}^2}\right)\right|^2.
\end{align}
Here, $G_{\rm F}$ is the Fermi constant, $\alpha_S$ and $\alpha_{\rm EM}$ define the strong and electromagnetic coupling constants, respectively. The function $\mathcal{P}$ can be defined as,
\begin{align}
  \mathcal{P}(x)=-\frac{[x-\psi(x)]}{x^2}\,,
 \end{align}
 where,
 \begin{align}
 \psi(x)&=\Bigg\{\begin{array}{cc}
 {\rm Arcsin}^2(\sqrt{x}), & x\leq 1\\
 -\frac{1}{4}\left[{\rm ln}\left(\frac{1+\sqrt{1-x^{-1}}}{1-\sqrt{1-x^{-1}}}\right)-i\pi\right]^2, & x>1\,.
 \end{array}
 \label{eq:glu_func}
\end{align}   
Thus, the proposed framework opens up 3-body CLFV decay routes of the form $\ell_A\to\ell_B\gamma\gamma$ and $\ell_A\to\ell_B gg$, which can possibly be used to check the validity of the available parameter space. However, the former decay channel, where the heavier lepton decays to a lighter one and two photons, is well-known in the literature and has also been explored through experiments, whereas the latter is mostly unconstrained and rarely discussed in theory. 

$\Gamma_\phi$ being extremely suppressed~$\left(\leq\mathcal{O}(10^{-22})~ {\rm GeV}\right)$, one can use the narrow width approximation to study the $\phi$-mediated 3-body CLFV decays~\cite{Cordero-Cid:2005vca}. Therefore, the corresponding branching ratios can be cast as,
\begin{align}
{\rm BR}(\ell_A\to\ell_B XX)\simeq {\rm BR}(\ell_A\to\ell_B\phi)\times {\rm BR}(\phi\to XX), 
\end{align}
where $X$ represents either photon or gluon. 
\subsection{$\pmb{\ell_A\to\ell_B\gamma\gamma}$}
\begin{figure}[!ht]
\centering
\subfloat[\qquad\qquad (a)]{\includegraphics[scale=0.65]{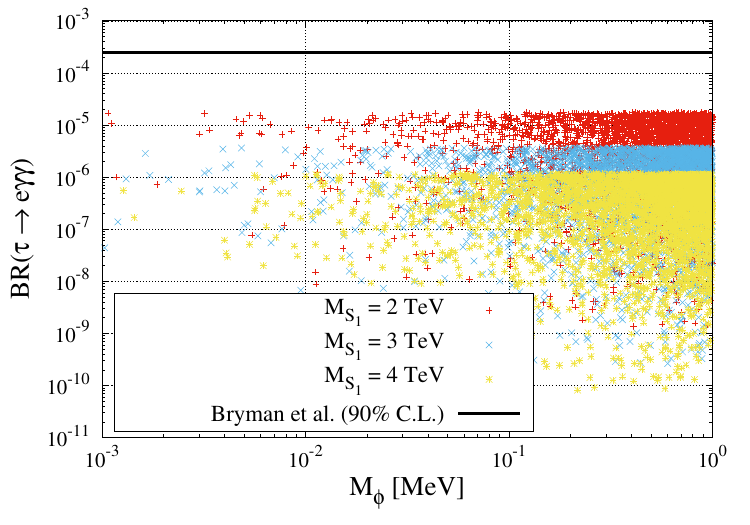}}\\
\subfloat[\qquad\qquad (b)]{\includegraphics[scale=0.65]{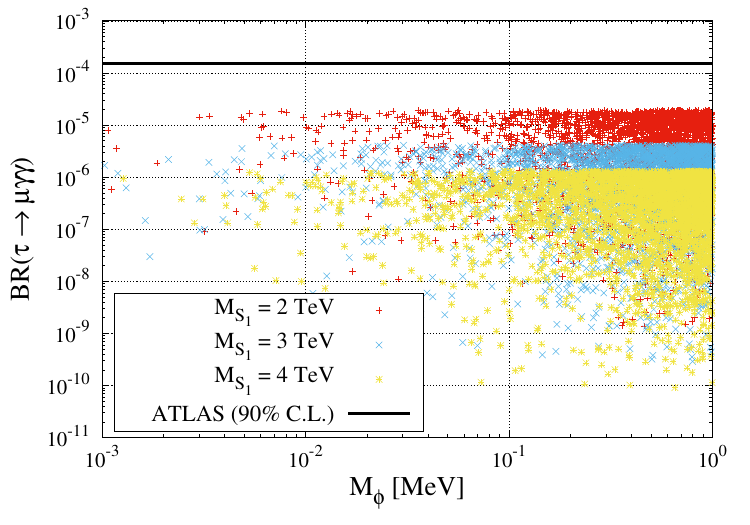}}\\
\subfloat[\qquad\qquad (c)]{\includegraphics[scale=0.65]{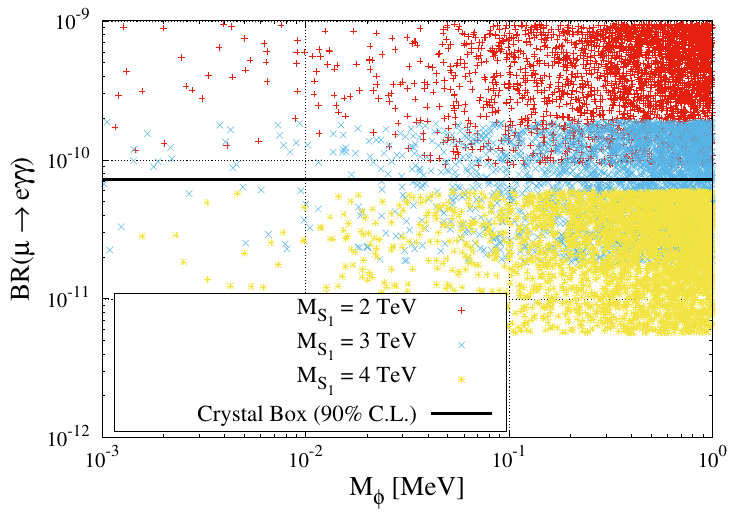}}
\caption{Variation of (a) BR$(\tau\to e\gamma\gamma)$, (b) BR$(\tau\to \mu\gamma\gamma)$, and  (c) BR$(\mu\to e\gamma\gamma)$ as a function of $M_\phi$ for $M_{S_1}=2$ TeV~(red), 3 TeV~(sky), and 4 TeV~(yellow). The black solid lines represent the corresponding experimental sensitivities~(at 90\% C.L.).}
\label{fig:BR_Y}
\end{figure}
This type of 3-body CLFV processes has been searched for a long time~(particularly for $\mu\to e\gamma\gamma$) through various experiments, and till now, no flavor violation has been observed. Thus, the null results have set certain upper limits on BR($\ell_A\to\ell_B\gamma\gamma$), which may further propagate into the $(\ell_A\to\ell_B\phi)$-allowed parameter space as more stringent constraints. Table~\ref{tab:LFV} presents the current experimental bounds on BR($\ell_A\to\ell_B\gamma\gamma$).
\begin{table}[!ht]
\centering
\begin{tabular}{|c|c|c|}
\hline
CLFV Observables & Upper Limits & Experiments\\
\hline\hline
BR($\mu\to e\gamma\gamma$) & $7.2\times 10^{-11}$ & Crystal Box~\cite{Bolton:1988af}\\
\hline
BR($\tau\to e\gamma\gamma$) & $2.5\times 10^{-4}$ & Bryman et al.~\cite{Bryman:2021ilc}\\
\hline
BR($\tau\to \mu\gamma\gamma$) & $1.5\times 10^{-4}$ & ATLAS~\cite{Angelozzi:2017oeg}\\
\hline
\end{tabular}
\caption{Experimental bounds on $\ell_A\to\ell_B\gamma\gamma$ processes.}
\label{tab:LFV}
\end{table} 
Clearly, the experimental bounds for BR$(\tau\to \ell_B\gamma\gamma)$~[$\ell_B=e,\,\mu$] are less stringent compared to those on $R_{\tau\ell_B}$ from Belle-II. It can be easily understood from  Fig.~\ref{fig:BR_Y}\,(a) and \ref{fig:BR_Y}\,(b) that the 3-body decay doesn't add any additional constraint in the $\tau$-sector, and effectively the entire parameter space is allowed with the existing upper limits. Therefore, Fig.~\ref{fig:R1} alone displays the actual permitted region where the production prospects of $\phi$ via $\tau$-decays can be studied.
Fig.~\ref{fig:BR_Y}\,(c) depicts the $M_\phi$ dependence of BR$(\mu\to e\gamma\gamma)$ for different values of $M_{S_1}$. The currently available experimental sensitivity restricts BR$(\mu\to e\gamma\gamma)<7.2\times 10^{-11}$~\cite{Bolton:1988af} and, thus, introduces the strongest bound on the considered parameter space. With all the $(\mu\to e\phi)$-specific NP couplings~(i.e., $|\Omega^{\mu e}_i|$) being fixed at their allowed values/ranges as listed in Table~\ref{tab:bound}, the results show that a sub-MeV SM-singlet scalar can only be produced through the flavor violating muon decay if $M_{S_1}> 2$ TeV. 
\subsection{$\pmb{\ell_A\to\ell_Bgg}$}
\begin{figure}[!ht]
\centering
\subfloat[\qquad\qquad (a)]{\includegraphics[scale=0.65]{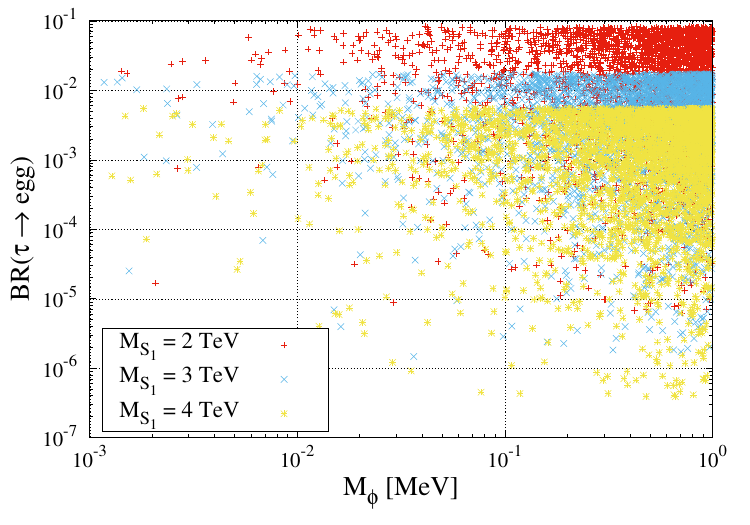}}\\
\subfloat[\qquad\qquad (b)]{\includegraphics[scale=0.65]{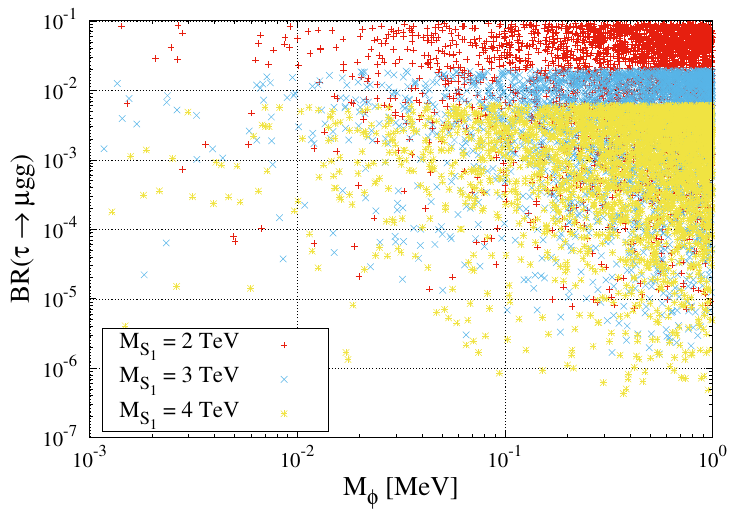}}\\
\subfloat[\qquad\qquad (c)]{\includegraphics[scale=0.65]{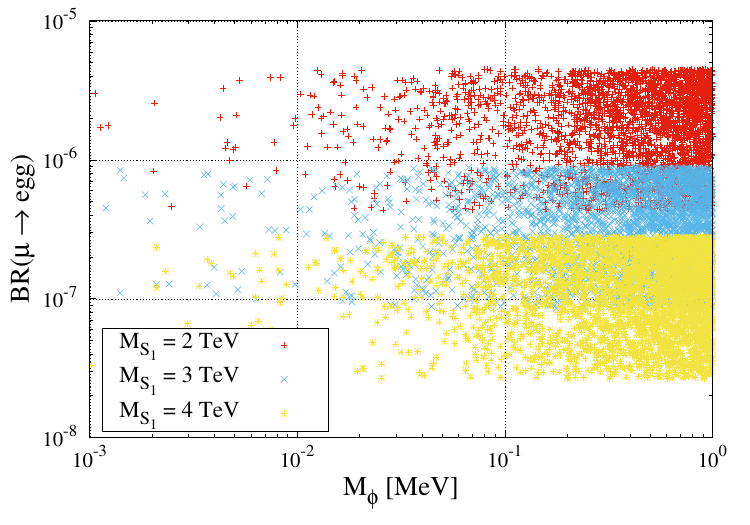}}
\caption{Variation of (a) BR$(\tau\to e\,gg)$, (b) BR$(\tau\to \mu\,gg)$, and (c) BR$(\mu\to e\,gg)$ as a function of $M_\phi$ for $M_{S_1}=2$ TeV~(red), 3 TeV~(sky), and 4 TeV~(yellow).}
\label{fig:BR_g}
\end{figure}
The proposed extension of SM can accommodate an exotic 3-body CLFV channel where the heavier charged leptons can decay to the lighter ones and two gluons. To the best of the author's knowledge, very little has been explored about this particular type of CLFV decays, as they can never be probed through low-energy experiments. Colliders using deep inelastic scattering can only be an option to search for $\ell_A\to\ell_Bgg$~\cite{AbdulKhalek:2022hcn}. For example, Refs.~\cite{Takeuchi:2017btl,Cirigliano:2021img} have discussed the gluonic contribution to the $\tau$-specific CLFV processes in an Electron-Ion/Proton Collider. 

Therefore, these decays might lead to completely new observables for probing the lepton flavor violation in the charged sector. Fig.~\ref{fig:BR_g} shows the model predictions for $\tau\to egg$, $\tau\to \mu gg$, and $\mu \to e gg$. Note that, ${\rm BR}(\ell_A\to\ell_B gg)/{\rm BR}(\ell_A\to\ell_B \gamma\gamma)\sim\mathcal{O}(10^3)$ as $\phi$ dominantly decays through the di-gluon channel. Therefore, in principle, $\ell_A\to\ell_B gg$ should have a better detection prospect compared to $\ell_A\to\ell_B \gamma\gamma$ if the background is substantially reduced~\cite{ATL-PHYS-PUB-2017-017, CMS-DP-2017-027, Andrews:2019faz}. Though, currently, there is no direct experimental bound on these processes, detectors looking for di-gluon signals from CLFV decays may test/falsify the present predictions in the future.
\section{Conclusion}
\label{sec:5}
\noindent
By the next decade, experiments searching for CLFV processes are going to play a vital role in discovering physics beyond the SM. The chances will definitely be enhanced with the increasing number of CLFV observables. Moreover, the observation will also be significant in understanding the possible NP theory associated with such a positive signal. This paper has extended the SM with a light SM-singlet scalar $\phi$ with mass $M_\phi<2m_e$ and a TeV-scale scalar LQ $S_1$. There is a wide range of theories where such a generic gauge-singlet scalar can be associated with a spontaneously broken global symmetry. The effective couplings of $\phi$ with the SM fields can be obtained at one-loop level in the presence of $S_1$. Phenomenologically, scalar LQs are well-motivated hypothetical particles that generate a large variety of CLFV processes. Thus, the non-observation of any CLFV process with the existing experimental sensitivities results in stringent constraints on the NP couplings at the quark-LQ-lepton vertices. Considering such a pre-constrained parameter space, the present paper has studied the discovery prospects of $\tau\to e\phi$, $\tau\to \mu\phi$, and $\mu\to e\phi$. Though for illustrations, 1 keV $\leq M_\phi\leq$ 1 MeV has been chosen, the results are equally valid for any generic scalar lighter than $2m_e$. $S_1$ being an EM-charged color-triplet scalar, $\phi$ can decay through di-photon and di-gluon channels. This induces the possibility of 3-body CLFV decays $\ell_A\to\ell_B\gamma\gamma$ and $\ell_A\to\ell_B gg$, where $\phi$ acts as the mediator. Though for the $\tau$-sector, the experimental bounds on $\tau\to\ell_B\gamma\gamma$~[$\ell_B=e,\,\mu$] are less stringent compared to those on the $\tau\to\ell_B \phi$, for $\mu$-sector BR$(\mu\to e\gamma\gamma)$ becomes crucial to constrain the parameter space. The most remarkable feature of this model is the possibility to predict CLFV observables corresponding to the 3-body decay $\ell_A\to\ell_B gg$, for all the possible channels. No direct experimental limit exists for such leptonic decays with two gluons in the final states. Using narrow-width approximation, the model has predicted for BR$(\tau\to e gg)$, BR$(\tau\to \mu gg)$, and BR$(\mu\to e gg)$ within the allowed parameter space. Therefore, a future experiment searching for $\ell_A\to\ell_B gg$ will be significant to test/constrain the model or may lead to a discovery.

%%%%%%%%%%%%%%%%%%%%%%%%%%%%%%%%%%%%%%%%%%%%%%%%%%%%%%%%%%%%%%%%%%%%%%%%%%%%%%%%%%%%%%    
\bigskip
\small \bibliography{Tau_CLFV}{}
\bibliographystyle{JHEPCust}    
    
\end{document}